\documentclass[12pt]{article}

\usepackage{amsthm,amsfonts,amsmath,graphicx}
\usepackage[sort&compress,numbers]{natbib}

\usepackage{amsthm}

\theoremstyle{plain}

\newtheorem{theorem}{Theorem}

\newtheorem{lemma}{Lemma}
\newtheorem{corollary}{Corollary}

\theoremstyle{definition}
\newtheorem{conjecture}{Conjecture}

\newtheorem{open}{Open Problem}

\newcommand{\twothmref}[2]{Theorems~\ref{thm:#1} and \ref{thm:#2}}

\newcommand{\thmref}[1]{Theorem~\ref{thm:#1}}

\newcommand{\figref}[1]{Figure~\ref{fig:#1}}

\newcommand{\corref}[1]{Corollary~\ref{cor:#1}}

\newcommand{\eqnref}[1]{\eqref{eqn:#1}}

\newcommand{\figlabel}[1]{\label{fig:#1}}

\newcommand{\thmlabel}[1]{\label{thm:#1}}
\newcommand{\lemlabel}[1]{\label{lem:#1}}

\newcommand{\corlabel}[1]{\label{cor:#1}}
\newcommand{\eqnlabel}[1]{\label{eqn:#1}}

\DeclareFontShape{OT1}{cmr}{b}{sc}
{<5><6><7><8><9><10><12><10.95><14.4><17.28><20.74><24.88>cmbcsc10}{}
\DeclareFontShape{OT1}{cmr}{bx}{sc}
{<5><6><7><8><9><10><12><10.95><14.4><17.28><20.74><24.88>cmbcsc10}{}

\DeclareFontShape{OT1}{cmr}{b}{tt}
{<5><6><7><8><9><10><12><10.95><14.4><17.28><20.74><24.88>cmbtt10}{}
\DeclareFontShape{OT1}{cmr}{bx}{tt}
{<5><6><7><8><9><10><12><10.95><14.4><17.28><20.74><24.88>cmbtt10}{}

\newcommand{\bracket}[1]{\ensuremath{\protect\left(#1\right)}}

\newcommand{\CEIL}[1]{\ensuremath{\protect\left\lceil#1\right\rceil}}
\newcommand{\FLOOR}[1]{\ensuremath{\protect\left\lfloor#1\right\rfloor}}
\newcommand{\floor}[1]{\ensuremath{\protect\lfloor#1\rfloor}}

\newcommand{\half}{\ensuremath{\protect\tfrac{1}{2}}}

\newlength{\marginboxwidth}

\usepackage{pifont}

\newcommand{\aaa}{\textup{(a)}}
\newcommand{\bbb}{\textup{(b)}}
\newcommand{\ccc}{\textup{(c)}}
\newcommand{\ddd}{\textup{(d)}}
\newcommand{\eee}{\textup{(e)}}

\title{\textbf{Characterisations of Intersection Graphs\\ by Vertex
Orderings}\,\thanks{Research supported by NSERC and COMBSTRU.}}

\author{David R. Wood\\[1ex]
\small School of Computer Science\\[-0.8ex]
\small Carleton University, Ottawa, Canada\\[0.5ex]
\small Department of Applied Mathematics\\[-0.8ex]
\small Charles University, Prague, Czech Republic\\[0.5ex]
\small \texttt{davidw@scs.carleton.ca}
}

\begin{document}
\maketitle

\begin{abstract}
Characterisations of interval graphs,  comparability graphs,
co-comparability graphs, permutation graphs, and split graphs in terms of
linear orderings of the vertex set are presented. As an application, it is
proved that interval graphs,  co-comparability graphs, AT-free graphs, and
split graphs have bandwidth bounded by their maximum degree. 
\end{abstract}

\section{Introduction}

We consider finite, simple and undirected graphs $G$ with vertex set $V(G)$,
edge set $E(G)$, and maximum degree $\Delta(G)$. The \emph{compliment} of $G$
is the graph $\overline{G}$ with vertex set $V(G)$ and edge set $\{vw:v,w\in
V(G),vw\not\in E(G)\}$. A \emph{vertex ordering} of $G$ is a total order
$(v_1,v_2,\dots,v_n)$ of $V(G)$. Let $\mathcal{S}$ be a finite family of sets. 
The \emph{intersection graph} of $\mathcal{S}$ has vertex set $\mathcal{S}$ and
edge set $\{AB:A,B\in S,A\cap B\ne\emptyset\}$.  This paper presents
characterisations of a number of popular intersection graphs in terms of vertex
orderings. 

In a vertex ordering $(v_1,v_2,\dots,v_n)$ of a graph $G$, the \emph{width} of
an edge $v_iv_j\in E(G)$ is $|i-j|$. The maximum width of an edge is the
\emph{width} of the ordering. The \emph{bandwidth} of $G$ is the minimum width
of a vertex ordering of $G$. Bandwidth is a ubiquitous concept with numerous
applications (see \citep{CCDG-Bandwidth-JGT82}). Obviously the bandwidth of $G$
is at least $\half\Delta(G)$. As an application of our results, we prove upper
bounds on the bandwidth of many intersection graphs $G$ in terms of
$\Delta(G)$. 

\section{Interval Graphs}

An \emph{interval graph} is the intersection graph of a finite set of closed
intervals in $\mathbb{R}$.  We have the following characterisation of interval
graphs.

\begin{theorem}
\thmlabel{Interval}
A graph $G$ is an interval graph if and only if $G$ has a vertex ordering
$(v_1,v_2,\dots,v_n)$ such that 
\begin{equation}
\eqnlabel{Interval}
\forall\,i<j<k,\;v_iv_k\in E(G)\;\Rightarrow\;v_iv_j\in E(G)\enspace.
\end{equation}
\end{theorem}

\begin{proof} Let  $(v_1,v_2,\dots,v_n)$ be a vertex ordering of $G$ satisfying
\eqnref{Interval}. For each vertex $v_i$, associate the interval $[i,r(i)]$,
where $v_{r(i)}$ is the rightmost neighbour of $v_i$. For every edge $v_iv_j\in
E(G)$ with $i<j$, $j\in[i,r(i)]\cap[j,r(j)]$.  For every non-edge $v_iv_j\in
E(\overline{G})$ with $i<j$, we have $r(i)<j$ by \eqnref{Interval}, and 
$[i,r(i)]\cap[j,r(j)]=\emptyset$.  Hence the intersection graph of
$\{[i,r(i)]:v_i\in V(G)\}$ is $G$.

Let $G$ be an interval graph. It is well known that we can assume that the
endpoints of the intervals are distinct. Let $(v_1,v_2,\dots,v_n)$ be a vertex
ordering of $G$ determined by increasing values of the left endpoints of the
intervals. For all $i<j<k$, the left endpoint of $v_j$ is between the left
endpoints of $v_i$ and $v_k$. If $v_iv_k\in E(G)$ then the right endpoint of
$v_i$ is to the right of the left endpoint of $v_k$. Thus the left endpoint of
$v_j$ is in the interval for $v_i$. Hence $v_iv_j\in E(G)$, as
claimed.
\end{proof}

As far as we are aware, \thmref{Interval} has not appeared in the literature,
although similar results are known. For example, \citet{GilmoreHoffman-CJM64}
proved that $G$ is an interval graph if and only if there is an ordering of the
maximal cliques of $G$ such that for each vertex $v$, the maximal cliques
containing $v$ appear consecutively.

\thmref{Interval} implies the following result of \citet{FominGolovach-DAM03}.

\begin{corollary}[\citep{FominGolovach-DAM03}]
Every interval graph $G$ has bandwidth at most $\Delta(G)$.
\end{corollary}

\begin{proof} In the vertex ordering  $(v_1,v_2,\dots,v_n)$ from
\thmref{Interval}, the width of an edge $v_iv_k\in E(G)$ is  $|\{v_iv_j\in
E(G):i<j\leq k\}|\leq\deg(v_i)\leq\Delta(G)$. \end{proof}


A \emph{proper interval graph} is the intersection graph of a finite set
$\mathcal{S}$ of closed intervals in $\mathbb{R}$ such that $A\not\subset B$
for all $A,B\in\mathcal{S}$. The following characterisation is due to
\citet{LoogesOlariu93} (see \citep{Corneil-DAM04}).

\begin{theorem}[\citep{LoogesOlariu93}]
\thmlabel{ProperInterval}
A graph $G$ is a proper interval graph if and only if $G$ has a vertex ordering
$(v_1,v_2,\dots,v_n)$ such that,
\begin{equation}
\eqnlabel{ProperInterval}
\forall\,i<j<k,\;v_iv_k\in E(G)\;\Rightarrow\;v_iv_j\in E(G)\;\wedge\;v_jv_k\in E(G)\enspace.
\end{equation}
\end{theorem}

It is easily seen that the bandwidth of a proper interval graph is one less
than the maximum clique size. Moreover, \citet{KaplanShamir-SJC96} proved that
the bandwidth of any graph $G$ equals the minimum, taken over all proper
interval supergraphs $G'$ of $G$, of the bandwidth of $G'$.

\section{Comparability Graphs}

Let $(P,\preceq)$ be a poset.  The \emph{comparability graph} of $(P,\preceq)$
has vertex set $P$, and distinct elements are adjacent if and only if they are
comparable under $\preceq$. We have the following characterisation of
comparability graphs. 

\begin{theorem}
\thmlabel{Comparability}
The following are equivalent for a graph $G$:\\
\aaa\ $G$ is a comparability graph,\\
\bbb\ $G$ has a vertex ordering $(v_1,v_2,\dots,v_n)$ such that,
\begin{equation}
\eqnlabel{Comparability}
\forall\,i<j<k,\;v_iv_j\in E(G)\;\wedge\;v_jv_k\in E(G)
\;\Rightarrow\;v_iv_k\in E(G)\enspace.
\end{equation}
\end{theorem}

\begin{proof} Let $G$ be the comparability graph of a poset $(V(G),\preceq)$.
A linear extension of $\preceq$ satisfies \eqnref{Comparability}. Given a
vertex ordering that satisfies \eqnref{Comparability},  define $v_i\prec v_j$
whenever $v_iv_j\in E(G)$ and $i<j$. Thus $(V(G),\preceq)$ is a poset, and $G$
is a comparability graph.
\end{proof}


A \emph{co-comparability graph} is a compliment of a comparability graph. As
illustrated in \figref{FunctionDiagram},  a \emph{function diagram} is a set
$\{c_i:1\leq i\leq n\}$, where for each $c_i$ is a curve $\{(x,f_i(x)):0\leq
x\leq 1\}$ for some function  $f_i:[0,1]\rightarrow\mathbb{R}$. If each $c_i$
is a line segment we say  $\{c_i:1\leq i\leq n\}$ is \emph{linear}.

\begin{figure}[hbt]
\vspace*{1ex}
\begin{center}\includegraphics{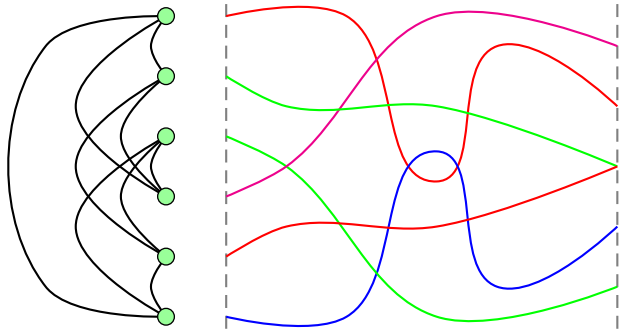}\end{center}
\vspace*{-2ex}
\caption{\figlabel{FunctionDiagram}A vertex ordering of the intersection graph of a function diagram.}
\end{figure}

\begin{theorem}
\thmlabel{CoComparability}
The following are equivalent for a graph $G$:\\
\aaa\ $G$ is a co-comparability graph,\\
\bbb\ $G$ is the intersection graph of a function diagram, and \\
\ccc\ $G$ has a vertex ordering $(v_1,v_2,\dots,v_n)$ such that,
\begin{equation}
\eqnlabel{CoComparability}
\forall\,i<j<k,\;
v_iv_k\in E(G)\;\Rightarrow\;v_iv_j\in E(G)\;\vee\;v_jv_k\in E(G)\enspace.
\end{equation}
\end{theorem}

\begin{proof} \citet{KGK86} and \citet{GRU-DM83} independently proved that
\aaa\ and \bbb\ are equivalent.

We now prove that \ccc\ implies \aaa. 
Suppose that $G$ has a vertex ordering
$(v_1,v_2,\dots,v_n)$ satisfying \eqnref{CoComparability}.
Define $v_i\prec v_j$ if $i<j$ and $v_iv_j\not\in E(G)$. 
Obviously $\prec$ is antisymmetric.
Suppose $v_i\prec v_j$ and $v_j\prec v_k$.
Then $i<k$ and $v_iv_k\not\in E(G)$, as otherwise 
\eqnref{CoComparability} fails.
That is, $v_i\prec v_k$. 
Thus $\prec$ is transitive, 
and $(V(G),\preceq)$ is a poset, whose
comparability graph is $\overline{G}$. 
Therefore $G$ is a co-comparability graph.

We now prove that \bbb\ implies \ccc.
Let $G$ be the intersection graph of a  function diagram 
$\{c_i:1\leq i\leq n\}$   with corresponding functions $\{f_i:1\leq i\leq n\}$.
Re-index so that  $f_i(0)\leq f_{i+1}(0)$ for all
$1\leq i\leq n-1$.  
Associate a vertex $v_i$ with each function $f_i$.   
Consider an edge $v_iv_k\in E(G)$ and a vertex $v_j$ with $i<j<k$.
There is a region $S$ bounded by $c_i$, $c_k$, and the line $X=0$, such that 
$c_j$ intersects the closed interior of $S$ and the closed exterior of $S$.
Thus $c_j$ intersects the boundary of $S$. Since $f_j$ is a function on
$[0,1]$, $c_j$ intersects the boundary of $S$ at a point on $c_i$ or $c_k$.
Thus $c_i\cap c_j\ne\emptyset$ or $c_j\cap c_k\ne\emptyset$.
Hence $v_iv_k\in E(G)$ or $v_jv_k\in E(G)$. 
That is, the vertex ordering $(v_1,v_2,\dots,v_n)$ satisfies
\eqnref{CoComparability}.
%
Note that we could have ordered the vertices with respect to  any fixed value
of $x_0\in[0,1]$, and in general, there are many vertex orderings that satisfy
\eqnref{CoComparability}.
\end{proof}

\begin{corollary}
\corlabel{CoComparabilityBandwidth}
Every co-comparability graph $G$ has bandwidth at most $2\Delta(G)-1$.
\end{corollary}

\begin{proof} In the vertex ordering $(v_1,v_2,\dots,v_n)$ from
\thmref{CoComparability}, the width of an edge $v_iv_k\in E(G)$ is at most
$|\{v_iv_j\in E(G):i<j<k\}|+|\{v_jv_k\in E(G):i<j<k\}|+1\leq
(\deg(v_i)-1)+(\deg(v_k)-1)+1\leq2\Delta(G)-1$. \end{proof}

It is interesting to ask whether \corref{CoComparabilityBandwidth} is
tight. It is easily seen that the complete bipartite graph $K_{n,n}$, which is
a co-comparability graph with maximum degree $n$, has bandwidth $3n/2$. 


\medskip Let $\pi$ be a permutation of $\{1,2,\dots,n\}$.  Let $\pi^{-1}(i)$
denote the position of $i$ in $\pi$. The \emph{permutation graph}  associated
with $\pi$ has vertex set $\{v_1,v_2,\dots,v_n\}$ and edge set
$\{v_iv_j:(i-j)(\pi^{-1}(i)-\pi^{-1}(j))<0\}$. The following characterisations of
permutation graphs can be derived from results of \citet{DushnikMiller41} and
\citet{BFR-Networks72}. Part (e) is proved as in
\twothmref{Comparability}{CoComparability}.

\begin{theorem}[\citep{DushnikMiller41,BFR-Networks72}]
\thmlabel{Permutation}
The following are equivalent for a graph $G$:\\
\aaa\ $G$ is a permutation graph,\\
\bbb\ $G$ is the intersection graph of a linear function diagram,\\
\ccc\ $G$ is a comparability graph and a co-comparability graph,\\
\ddd\ $G$ is the comparability graph of a two-dimensional poset,\\
\eee\ $G$ has a vertex ordering that simultaneously 
satisfies \eqnref{Comparability} and \eqnref{CoComparability}.
\end{theorem}

\section{AT-free Graphs}

An \emph{asteroidal triple} in a graph consists of an independent set of three
vertices such that each pair is joined by a path that avoids the neighbourhood
of the third. A graph is \emph{asteroidal triple-free} (\emph{AT-free}) if it
contains no asteroidal triples. 


\begin{lemma}
Every AT-free graph $G$ has bandwidth at most $3\Delta(G)$.
\end{lemma}

\begin{proof} A \emph{caterpillar} is a tree for which a path (called the
spine) is obtained by deleting all the leaves. Let $(v_1,v_2,\dots,v_m)$ be the
spine of a caterpillar $T$. The vertex ordering of $T$ obtained by inserting
the leaves adjacent to each $v_i$ immediately after $v_i$ has  bandwidth at
most $\Delta(T)$. 

\citet{KKM-JAlg99} proved that every (connected) AT-free graph $G$ has a
spanning caterpillar subgraph $T$, and adjacent vertices in $G$ are at
distance at most four in $T$. Moreover, for any edge $vw\in E(G)$ with $v$ and
$w$ at distance four in $T$, both $v$ and $w$ are leaves of $T$. Consider the
above vertex ordering of $T$ to be a vertex ordering of $G$. The bandwidth is
at most $3\Delta(T)\leq 3\Delta(G)$. \end{proof}

\section{Chordal Graphs}

A \emph{chord} of a cycle $C$ is an edge not in $C$ connecting two vertices in
$C$. A graph is \emph{chordal} if every induced cycle on at least four vertices
has at least one chord. The following famous characterisation of chordal graphs
is due to \citet{Dirac61}, \citet{FG-PJM65}, and \citet{Rose-JMAA70}.

\begin{theorem}[\citep{Rose-JMAA70,Dirac61,FG-PJM65}]\thmlabel{Chordal}
The following are equivalent for a graph $G$:\\
\aaa\ $G$ is a chordal,\\
\bbb\ $G$ is the intersection graph of subtrees of a tree, and\\
\ccc\ $G$ has a vertex ordering $(v_1,v_2,\dots,v_n)$ such that,
\begin{equation}
\eqnlabel{Chordal}
\forall\,i<j<k,\;
v_iv_j\in E(G)\;\wedge\;v_iv_k\in E(G)\;\Rightarrow\;v_jv_k\in E(G)\enspace.
\end{equation}
\end{theorem}

A striking generalisation of \thmref{Chordal} for $k$-chordal graphs is given
by \citet{DKT-TCS97}. A vertex ordering satisfying \eqnref{Chordal} is called a
\emph{perfect elimination} vertex ordering. It is not possible to bound the
bandwidth of every chordal graph $G$ in terms of $\Delta(G)$. For example, the
bandwidth of the complete binary tree on $n$ vertices is $\approx n/\log n$
\citep{Smithline-DM95}.


\medskip A graph $G$ is a \emph{split graph} if $V(G)=K\cup I$, where $K$
induces a complete graph of $G$, and $I$ is an independent set of $G$.

\begin{theorem}\thmlabel{Split}
The following are equivalent for a graph $G$:\\
\aaa\ $G$ is a split graph,\\
\bbb\ $G$ is chordal and $\overline{G}$ is chordal,\\
\ccc\ $G$ has a vertex ordering $(v_1,v_2,\dots,v_n)$ simultaneously satisfying
\eqnref{Chordal} and
\begin{equation}
\eqnlabel{Split}
\forall\,i<j<k,\;
v_iv_j\in E(G)\;\Rightarrow\;v_jv_k\in E(G)\;\vee\;v_iv_k\in E(G)\enspace,
\end{equation}
\ddd\ $G$ has a vertex ordering $(v_1,v_2,\dots,v_n)$ such that,
\begin{equation}
\eqnlabel{SimpleSplit}
\forall\,i<j<k,\;
v_iv_j\in E(G)\;\Rightarrow\;v_jv_k\in E(G)\enspace.
\end{equation}
\end{theorem}

\begin{proof}
\citet{FH77} proved that \aaa\ and \bbb\ are equivalent.

Observe that \ddd\ implies \ccc\ trivially. 
We now prove that \aaa\ implies \ddd. 
Let $G$ be a split graph with $V(G)=K\cup I$, where $K$ induces a complete
subgraph and $I$ is an independent set. Let $m=|I|$.
Consider a vertex ordering $(v_1,v_2,\dots,v_n)$ of $G$ where
$I=\{v_1,v_2,\dots,v_m\}$ and $K=\{v_{m+1},v_{m+2},\dots,v_n\}$. 
Suppose that $1\leq i<j<k\leq n$ and $v_iv_j\in E(G)$. 
There is no edge with both endpoints in $I$.
Thus $j\geq m+1$, and both $v_j,v_k\in K$.
Hence $v_jv_k\in E(G)$, and 
$(v_1,v_2,\dots,v_n)$ satisfies \eqnref{SimpleSplit}.

It remains to prove that \ccc\ implies \bbb.
Let $(v_1,v_2,\dots,v_n)$ be a vertex ordering of a graph $G$ 
satisfying \eqnref{Chordal} and \eqnref{Split}.
By \thmref{Chordal}, $G$ is chordal.
Equation \eqnref{Split} is equivalent to:
\begin{equation}
\eqnlabel{SplitDual}
\forall\,i<j<k,\;
v_jv_k\in E(\overline{G})\;\wedge\;v_iv_k\in E(\overline{G})
\;\Rightarrow\;
v_iv_j\in E(\overline{G})
\enspace.
\end{equation}
That is, $(v_n,v_{n-1},\dots,v_1)$ is a perfect elimination vertex ordering of
$\overline{G}$. By \thmref{Chordal}, $\overline{G}$ is chordal.
\end{proof}

\begin{lemma}
\lemlabel{SplitBandwidth}
Every split graph $G$ has bandwidth at most $\Delta(G)\,(\Delta(G)+2)$. For all
$\Delta\geq2$ there is a split graph $G$ with $\Delta(G)=\Delta$, and $G$
has bandwidth at least $\Delta(G)^2/12$.
\end{lemma}

\begin{proof} First we prove the upper bound. 
Let $G$ be a split graph with $V(G)=K\cup I$, where $K$
induces a complete subgraph, and $I$ is an independent set. 
The result is trivial if $K=\emptyset$. 
Now assume that $K\ne\emptyset$. 
Let $I_0$ be the set of isolated vertices in $G$.
Consider a vertex ordering $\pi$ 
in which the vertices in $I_0$ precede all other vertices.
Let $I_1=I\setminus I_0$.
Regardless of the order of $I_1\cup K$, the bandwidth of $\pi$ is at most
$|I_1|+|K|-1$. Thus it suffices to prove that
$|I_1|+|K|\leq\Delta(G)\,(\Delta(G)+2)+1$.

If $I_1=\emptyset$ then $\pi$ has bandwidth $\Delta(G)$. 
Now assume that $I_1\ne\emptyset$. 
Let $a$ be the average degree of vertices in $I_1$. 
Thus $1\leq a\leq |K|$. 
For each vertex $v\in K$, let  $b_v=\deg(v)-|K|+1$. 
That is, $b_v$ is the number of edges between $v$ and $I_1$. 
Let $b=\sum_{v\in K}b_v/|K|$. Thus $a|I_1|=b|K|$, which
implies that 
\begin{equation*}
|I_1|+|K|
\;=\;
\frac{b|K|}{a}+|K|
\;=\;
\frac{(b+a)|K|}{a}\enspace.
\end{equation*}
Now $\Delta(G)$ is at least the average degree of the vertices in $K$.
That is, $\Delta(G)\geq |K|-1+b$. Hence
\begin{equation*}
\frac{|I_1|+|K|}{(\Delta(G)+1)^2}
\;\leq\;
\frac{(b+a)|K|}{a(|K|+b)^2}\enspace.
\end{equation*}
Since $a\leq |K|$,
\begin{equation*}
\frac{|I_1|+|K|}{(\Delta(G)+1)^2}
\;\leq\;
\frac{|K|}{a(|K|+b)}
\enspace.
\end{equation*}
Since $a\geq1$ and $b\geq0$, 
\begin{equation*}
|I_1|+|K|
\;\leq\;
(\Delta(G)+1)^2
\;=\;\Delta(G)\,(\Delta(G)+2)+1\enspace,
\end{equation*}
as required.

Now we prove the lower bound. 
Given $\Delta$, let $n=\floor{\Delta/2}$. 
Let $G$ be the split graph with $V(G)=K\cup I$, where 
$K$ is a complete graph on $n$ vertices, and
$I$ is an independent set on $n(\Delta-n+1)$ vertices, 
such that every vertex in $K$ is adjacent to $\Delta-n+1$ vertices in $I$, 
and every vertex in $I$ is adjacent to exactly one vertex in $K$.
Clearly $G$ has diameter 3, maximum degree $\Delta$, 
and $n+n(\Delta-n+1)=n(\Delta+n-2)$ vertices. 
It is easily seen that every connected graph with $n'$ vertices and 
diameter $d'$ has bandwidth at least
$(n'-1)/d'$ \citep{CCDG-Bandwidth-JGT82}. 
Thus $G$ has bandwidth at least 
\begin{equation*}
\frac{1}{3}\bracket{ n(\Delta-n+2)-1 }
\;=\;
\frac{1}{3}\bracket{ \FLOOR{\frac{\Delta}{2}}\CEIL{\frac{\Delta+2}{2}}-1 }
\;\geq\;
\frac{\Delta^2}{12}\enspace.
\end{equation*}
\end{proof}


\bibliographystyle{myBibliographyStyle}
\bibliography{myBibliography,myConferences}

\end{document}